%% file: manuscript.tex
\documentclass{article}
\usepackage{PRIMEarxiv}
\usepackage[utf8]{inputenc} 
\usepackage[T1]{fontenc}    
\usepackage{booktabs} 
\usepackage{multirow}
\usepackage{siunitx}
\usepackage{subcaption}
\usepackage[most]{tcolorbox}
\usepackage[export]{adjustbox}
\usepackage{hyperref}
\usepackage{algorithm, algpseudocode}
\usepackage{hyperref}
\usepackage{placeins}
\usepackage{amsmath,amssymb,amsfonts}
\usepackage{graphicx}
\usepackage{acro}
\input{acronyms}
\def\BibTeX{{\rm B\kern-.05em{\sc i\kern-.025em b}\kern-.08em
    T\kern-.1667em\lower.7ex\hbox{E}\kern-.125emX}}
\usepackage{url}
\usepackage{hyperref}
\usepackage[none]{hyphenat}

\title{Quality Assurance in MLOps Setting: An Industrial Perspective}

\author{
  Ayan Chatterjee \\
  Dept of Mathematics and Computer Science, Karlstad University, 651 88 Karlstad, Sweden \\
  \texttt{ayan.chatterjee@kau.se} \\
   \And
  Bestoun S. Ahmed \\
  Dept of Mathematics and Computer Science, Karlstad University, 651 88 Karlstad, Sweden\\
  Dept of Computer Science, FEE, Czech Technical University in Prague, Czechia\\
  \texttt{bestoun@kau.se} \\
  \And
  Erik Hallin \\
  Uddeholms AB, Uvedsv{\"a}gen, Hagfors, 683 33, V{\"a}rmlands l{\"a}n, Sweden\\
  \texttt{erik.hallin@uddeholm.com} \\
  \And
  Anton Engman \\
  Uddeholms AB, Uvedsv{\"a}gen, Hagfors, 683 33, V{\"a}rmlands l{\"a}n, Sweden\\
  \texttt{anton.engman@uddeholm.com} \\
}

\begin{document}
\maketitle

\begin{abstract}
Today, machine learning (ML) is widely used in industry to provide the core functionality of production systems. However, it is practically always used in production systems as part of a larger end-to-end software system that is made up of several other components in addition to the ML model. Due to production demand and time constraints, automated software engineering practices are highly applicable. The increased use of automated ML software engineering practices in industries such as manufacturing and utilities requires an automated Quality Assurance (QA) approach as an integral part of ML software. Here, QA helps reduce risk by offering an objective perspective on the software task. Although conventional software engineering has automated tools for QA data analysis for data-driven ML, the use of QA practices for ML in operation (MLOps) is lacking. This paper examines the QA challenges that arise in industrial MLOps and conceptualizes modular strategies to deal with data integrity and Data Quality (DQ). The paper is accompanied by real industrial use-cases from industrial partners. The paper also presents several challenges that may serve as a basis for future studies.
\end{abstract}


\section{Introduction}

\begin{figure*}
    \centering
    \includegraphics[scale=0.45]{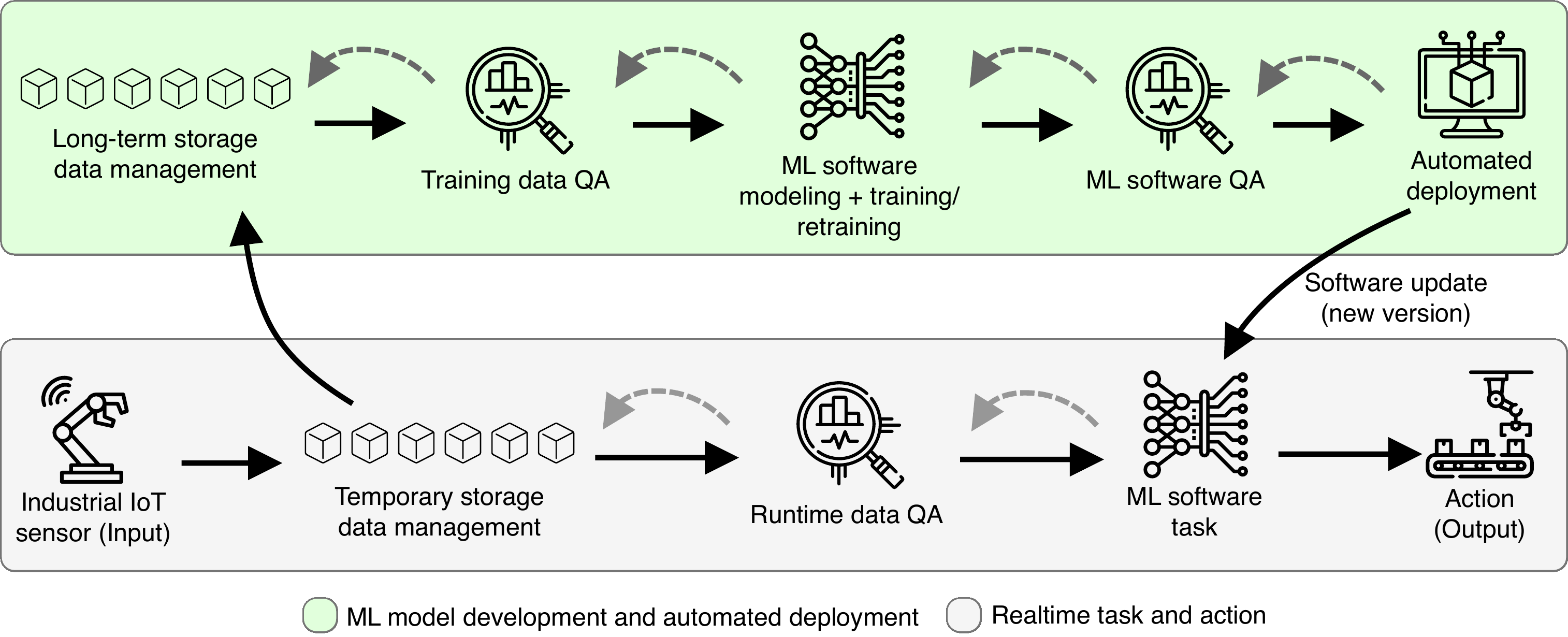}
    \caption{A high-level infographic overview of QA in MLOps architecture, inspired from AWS \cite{AWSMLOps} and Azure \cite{MSMLOps} architectures. (The straight lines in the figure represent data flow and the dashed lines represent feedback.)}
    \label{fig:mlopsFlow}
\end{figure*}

Traditionally, the software development life cycle includes quality assurance (QA) practices and automated tools to assess the quality of the software system. Due to the increasing functionalities of recent developments in cyber-physical systems in Industry 4.0 and human-centered development in Industry 5.0, ML programs that have core functionality are becoming increasingly ML software systems. ML software development in operation is an ongoing development in automated software engineering that sets up automated pipelines for continuous training and deployment of ML software. Recent advances toward a continuous development cycle of ML software are known as MLOps \cite{mlops2021} or AIOps \cite{aiops2019}. An MLOps pipeline, for example, in \ac{aws} \cite{AWSMLOps} and Microsoft Azure \cite{MSMLOps}, generally consists of ML software development and automated deployment and monitoring (or Ops). Then, trained ML software is deployed for real-time prediction and classification tasks and an action. Inspired by such architectures, we conceptualize an infographic diagram of the overview of the MLOps architecture in Figure \ref{fig:mlopsFlow}, which shows the continuous and automated development of ML software. Such ML software, which does not follow the conventional way of authoring software code, is a black box \cite{blackbox2021} and is data-driven \cite{datadriven2019}. Thus, shifting the requirements for QA, which previously relied on exposure to software code for Software Quality Assurance (SQA) and Data Quality (DQ) professionals and tools for data QA \cite{Lee2003}.

Data and SQA ensure compliance with the level of accuracy, data security, and performance scripted in an organization's QA policies and enforced by the SQA plan in classical software development in operation (DevOps) \cite{sqa2022}. Such a QA plan in MLOps requires robust software testing approaches and data QA in a software development pipeline. Recent developments in SQA of ML are enabled by methods such as domain-specific software testing \cite{drivingTest2021} and mutation testing \cite{mutation2021} that attempt to detect faults in ML software by systematically modifying the test data. When it comes to data QA, data integrity, trustworthiness, and DQ, they are measured by individual data dimensions such as timeliness, uniqueness, relevancy, and location intelligence \cite{dataQAbook2021}. However, the dimensions for a given software pipeline are pre-determined for a given application, and we discovered a lack of implementation for industrial MLOps in our search.

This paper attempts to address the following research questions (RQs):
\begin{itemize}
    \item \textbf{RQ 1: What are the industrial QA challenges when conventional software testing and QA are not viable for MLOps?}\\
    We collaborated with partners in the manufacturing and utilities sector to discover and outline the QA challenges in industrial MLOps for real \ac{uc}s.
    \item \textbf{RQ 2: What approach can we take when dealing with QA in industrial MLOps?}\\
    We propose concept for a modular approach to data QA, where each component of data QA is an ML software in the industrial MLOps architecture mentioned in Figure \ref{fig:mlopsFlow}, with automated training, selection and deployment of data dimensions. Furthermore, we conceptualize a practical and applicable MLOps architecture to enable QA and the continuous delivery of ML software.
\end{itemize}

\section{MLOps software architecture and QA challenges} \label{sec:architecture}

\subsection{Architecture}
Network architecture has undergone significant advances in recent years \cite{mlopsArc2022, crossPlatformEff2022}, making it possible for modern industries and factories of the future to offer a backbone for MLOps setup, minimize latency, and enable time-sensitive hard and soft deadline tasks. As a result, we have identified three network architectures, each of which depends on how far \ac{iot} data travel for ML software tasks. The architecture categories are as follows:
\begin{itemize}
    \item \textbf{End-to-end on-device processing:} The first is an end-to-end software pipeline where data acquisition, QA, ML software processing and all actions occur on the same device. These are relatively lightweight, usually a single action tiny ML task. Even if the ML software, in this case, is initially delivered externally, adaptation and retraining of the model occur on the fly. Due to the lack of processing power and devices running on a battery, QA is challenging in this situation.
    \item \textbf{Data processing on edge and fog nodes:} In contrast to the previous design, this architecture is built on local networks to handle computational operations on edge or fog nodes. Although this architecture has increased latency compared to on-device processing, this has the advantage of performing data analytics and action using information from multiple sources in the network.
    \item \textbf{IoT-edge/fog-cloud architecture:} The third architecture utilizes cloud services such as Amazon Web Services, Google Cloud, and Microsoft Azure, in addition to processing on-device, edge, and fog. This architecture has the highest latency among others, but it has the ability to execute collective analytics from all connected devices and historical data and is not resource constrained. Additionally, this enables centralized, decentralized, and federated ML applications.
\end{itemize}

Although the architecture categories differ for each application, they combine to form a hybrid network architecture for a suite of automated ML applications. Furthermore, because the hardware resources and time constraints available on the IoT device, the edge or fog network, and the cloud are vastly different, this creates a challenge for QA for industrial MLOps.

\subsection{QA Challenges}

Working in collaboration with industrial partners, we have identified the following challenges in relation to QA that are applicable to any architecture:

\begin{itemize}
    \item \textbf{Modelling challenges:} External factors such as anomaly and noise deviate from the sensor data or signals from its ideal measurements. Furthermore, QA for ML must automatically accommodate class imbalance and drifts \cite{drift2022}, where the current instance of the containerized ML application has observed changes in the test data compared to the data on which it was trained. There is a lack of robust AutoML strategies and algorithms for such external factors in automated QA.
    \item \textbf{Resource, time, and scalability:} While some industrial processes are sparse, others are frequent enough to produce big data. For big data, it is challenging to perform the necessary data QA stages in a timely manner. In addition, industries scale up frequently, and system integration of new hardware may differ from existing machinery. The rapid adoption of new hardware in existing QA processes poses a scalability issue.
    \item \textbf{Architectural constraints:} Industrial MLOps architectures have network components, each with its capabilities and limitations. Network components, such as routers and switches, have capabilities such as the number of data packets they can handle at a given time, network scheduling, and routing. Additionally, these components are subject to network attacks, such as distributed denial of service, for example \cite{industrialReq2021}. This contributes to the total latency that a data packet needs to traverse from IoT sensors to QA microservices running on edge/fog and is a challenge for a robust and scalable architecture.
    \item \textbf{Lack of production data in manufacturing processes:} Some manufacturing operations are time-intensive or infrequent/sparse, typically occurring on average once or twice a day. One such example is electroslag remelting in the steel industries \cite{ESROverview2016}. This indicates that such operations will have a few hundred manufacturing events over the course of a year, and the lack of data is a challenge for the ML software. Although data augmentation or a principled approach to generating synthetic data has been used to fill the gaps, it poses a QA challenge to strategically split the limited ground truth data for robust training and testing \cite{ESRTesting2022}.
    \item \textbf{Compliance with regulatory, export control, and ISO standards:} Data is often subject to regulatory requirements from government entities. For example, with medical equipment that requires additional regulatory requirements for increased safety, user data is subject to the General Data Protection Regulation (GDPR\footnote{https://gdpr-info.eu/}) in Europe and defense data are subject to export control. These regulations are local to a region and are not universally applicable. In automated and continuous ML software development, adhering to such regional regulations is challenging for automated QA.
\end{itemize}

\section{Modular QA for industrial MLOps} \label{sec:prop}
To accommodate both lightweight and cloud-based ML software pipelines for industrial applications, we designed a modular QA solution. The modular architecture is predicated on the notion that every data QA step needs its unique collection of dimensions. And the set of data QA dimensions for each data QA step is determined by the answer to the following question in an organization's SQA plan: \textit{What are the automated steps taken as a result of data QA?}

\subsection{ML software architecture}

Knowledge of automated steps or actions serves as the foundation for QA strategy. To make this behavior possible, we divided the overall architecture of QA for ML into three phases: (i) definition and formulation, (ii) dimension selection, and (iii) QA model training, each answering the following questions:
\begin{itemize}
    \item \textbf{Definition and formulation:} \textit{How does an organization define and formulate QA dimensions such as trustworthiness, relevancy, and privacy?}
    \item \textbf{Dimension selection:} \textit{What minimum QA dimensions are necessary to achieve the QA actions?}
    \item \textbf{QA model training:} \textit{How the selected QA dimensions can be trained and weighed automatically in industrial MLOps?}
\end{itemize}

\subsubsection{Definition and formulation}
In the first stage, given the objectives and QA policies of an organization, formal definitions and mathematical formulations need to be developed for each QA dimension. While the rest of the QA for the ML strategy is automated, all QA dimensions require a definition. The definition changes as the organization's policy shifts. Although some QA dimensions are universal, others depend on the objectives of a company. Sensor noise, for example, is an ubiquitous component that affects all streaming data from industrial IoT sensors. Other QA indicators, such as contextual DQ and data integrity depend on the QA strategy of the organization. Furthermore, for MLOps, it is also important to define QA actions. For example, a QA action is to identify whether the streaming data is relatively clean of noise and anomalies with intrinsic DQ metrics and applicable to current ML software in production with contextual DQ metrics.

\subsubsection{Dimension selection}

Following the definition of all dimensions of QA, the next step is to determine which dimensions are relevant to a given set of QA actions. The QA dimensions are then arranged in a connected graph structure with parent-child relationships. For example, in the literature, these parent-child relationships are structured as follows: data integrity (parent) with DQ, data integration, data enrichment, and location intelligence (children). Data completeness, timeliness, and relevancy (children) are aspects of DQ (parent), further classified into intrinsic and contextual subcategories. In QA for ML, the selection of dimensions is achieved by first calculating all dimensions of DQ from the test data and then minimizing the number of child nodes required. We have combined dimension selection and model training, which can be done iteratively as one process.

\subsubsection{QA model training}

\begin{figure}[H]
    \centering
    \includegraphics[scale=0.55]{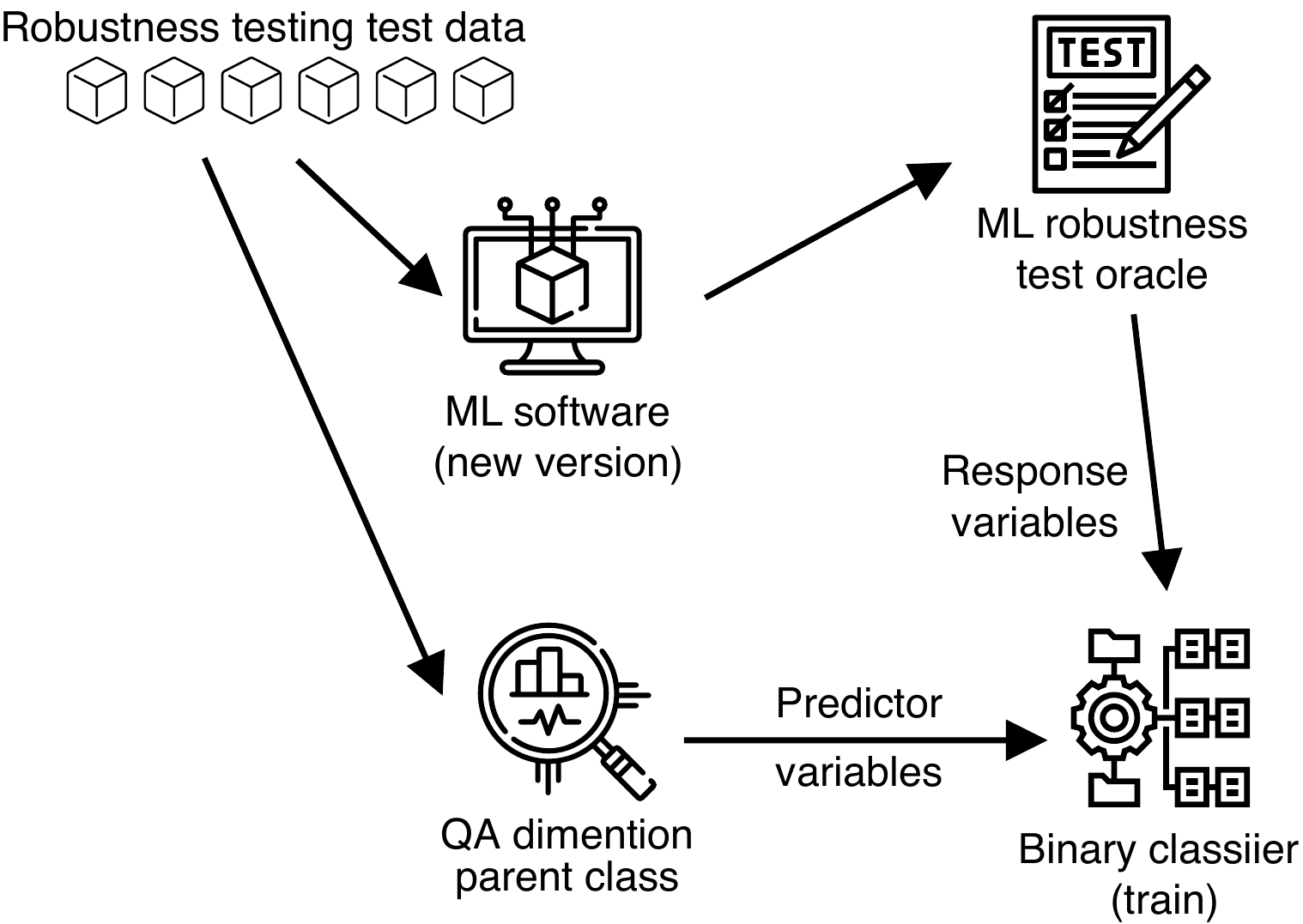}
    \caption{Robustness testing of each parent class in the model training phase.}
    \label{fig:classificationParent}
\end{figure}

The final step before deployment is QA for the ML model training. An approach to this problem is to iteratively perform dimension selection and model training, where the QA for ML model is trained with dimensions fixed, and dimension selection is performed to promote sparsity in the QA for ML graph while keeping the model fixed. Both steps are performed iteratively until the QA for ML model converges. For each parent in the model training process, robustness testing, a software testing methodology that examines the boundary conditions of a software \cite{MLTesting}, is carried out with the test data for a binary pass/fail score, as shown in Figure \ref{fig:classificationParent}.

The outcomes of each parent's robustness testing are then considered for the desired automated actions, allowing action-based QA for ML. For example, a score of "pass" for "intrinsic DQ" and "fail" for "contextual DQ" means that the streaming data are relatively clean and usable without the need of data cleaning (i.e., less noise, anomaly-free, and without NaN/missing values) but is not relevant for the current ML software in production. Still, it might be relevant for other ML software. In this case, with QA actions, the streaming data are set up to be sent to the QA for ML strategy in a different software pipeline or offline storage for future study.

\subsection{Real-time QA classification and action}

After training and testing, the next stage is to deploy QA for ML model using automated software deployment methods. Docker\footnote{https://www.docker.com/} and kubernetes\footnote{https://kubernetes.io/} are common tools for the deployment of ML software in a containerized manner. One such roadmap is to use Python scripts for model training and then generate a Docker container. The Docker container is then pushed to the Docker Hub, where it is then pulled to the Kubernetes edge nodes for real-time processing of the streaming data and action.

\begin{figure*}
    \centering
    \includegraphics[scale=0.55]{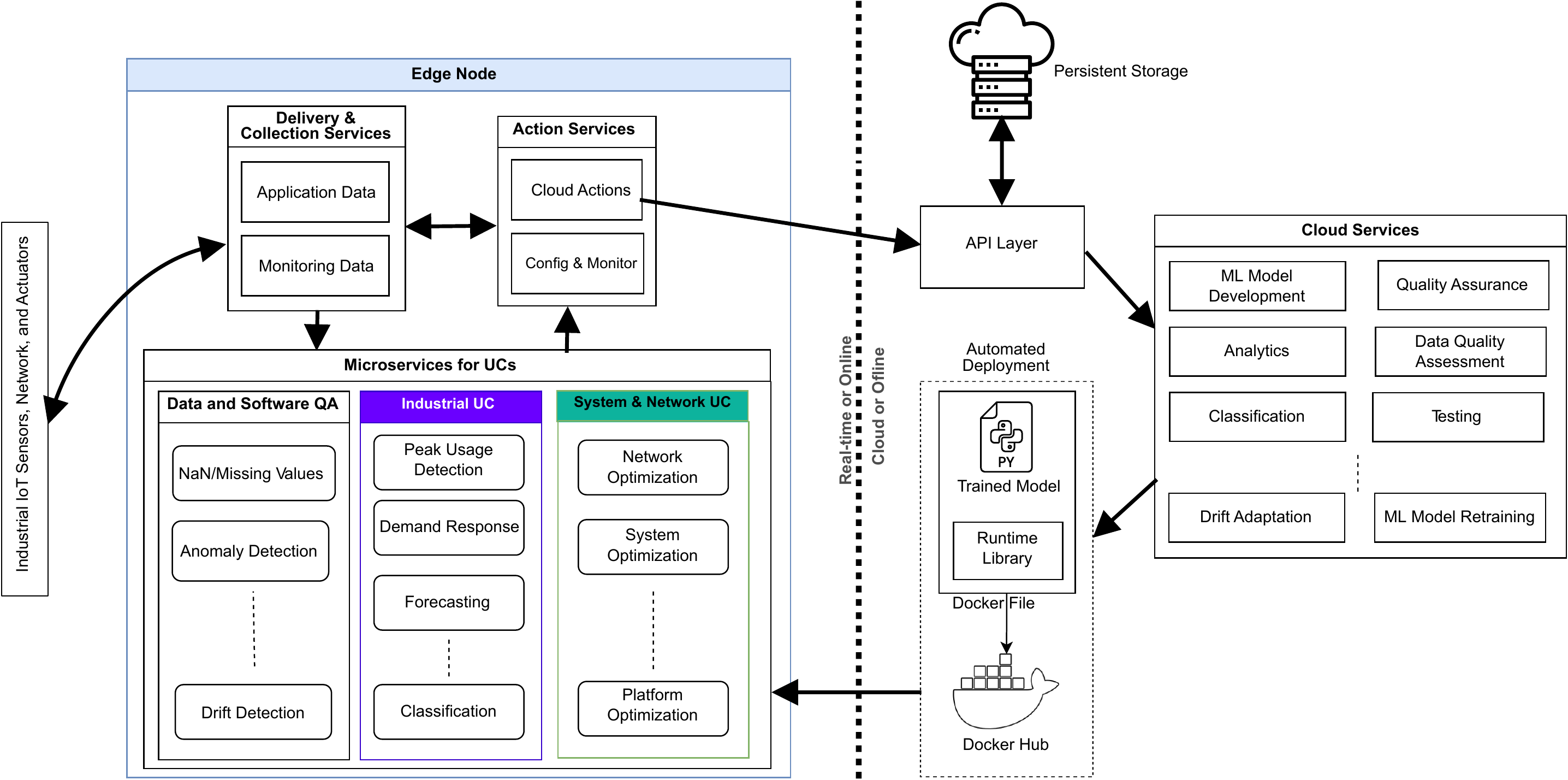}
    \caption{An architectural view of the proposed MLOps utilizing IoT, edge, and cloud for continuous ML software and QA for ML model training, testing, deployment, and real-time analytics and action.}
    \label{fig:edgeArc}
\end{figure*}

To support industrial \ac{uc}s and QA in MLOps, we designed an applicable and practical IoT-edge-cloud architecture for the continuous delivery of ML software. \ac{uc}s in industrial MLOps require support for two types of data transmission across the network: application and monitoring data. The incoming streams of IoT sensor data necessary for routine industrial applications are known as application data. The network data reflect the monitoring information, such as CPU and memory use in real time. To accommodate both types of data in the architecture, we designed the architecture in figure \ref{fig:edgeArc} that handles both real-time and offline data by combining microservices on the edge node with cloud services for ML training and offline applications. The architecture supports the following:
\begin{itemize}
    \item \textbf{Automated ML model retraining}: ML software, including AutoML software, once trained in `cloud services' is automatically deployed to the edge as an Edge AI software on the `edge node'. Once deployed for the first time, data flows from the network to the delivery and collection microservices to the cloud API layer. Once model degradation has occurred, the ML software is re-trained or adapted in the cloud. A new version of the ML software is automatically deployed on the edge node.
    \item \textbf{Real-time and automated decision and action}: Data transfer from the network to the delivery and collection microservices to the microservices for \ac{uc}s. Microservices for \ac{uc}s are then sent to Action Services to execute the appropriate action. The action signal is sent back to the network.
    \item \textbf{Automated QA assessment}: Similar to real-time and automated decision and action, data flows from the network to microservice delivery and collection to microservices for \ac{uc}s and Action Services. The microservices for \ac{uc}s here are containers that evaluate the quality of the incoming data. Furthermore, the quality of the containerized ML software for model degradation. Further, the action signal from the action microservice is sent to the Cloud API layer for the execution of the appropriate cloud service(s).
    \item \textbf{Message forwarding for cloud services}: In this scenario, for long-term storage, the message or data are forwarded from the network to the delivery and collection microservice to the cloud API layer for persistent storage.
\end{itemize}

\section{Future directions} \label{sec:usecase}
The modular QA and the MLOps architecture collectively enable real-time execution of real-time ML and action supported by data and SQA processes. However, research in this area is still in its infancy, and we have identified the following vectors for future direction:

\begin{itemize}
    \item \textbf{Software testing:} The principles of Chaos Engineering state that systems react differently based on surroundings and traffic patterns. This is true for software, and automated software engineering and software testing procedures, such as shadow testing, allow the software to be analyzed in terms of how it performs in different environments. The log data of the software in various environments is then analyzed. The use of SQA opens the possibility of potential research to automate shadow testing and other automated software tests. An industrial application where automated software testing with QA is useful is in compressed air. Compressing air is an energy-intensive process that is used in industries to remove dust and cooling \cite{benedetti2018}. In addition to wasting air/gas and increasing operational expenses, leaks can also be a point of entry for contaminants to enter the system. Therefore, the process must be continuously monitored for leaks to reduce overall energy and operational cost. Streaming data in compressed air are often subject to anomaly and noisy measurements. Monitoring and predicting future air pressure readings based on historical data is one such \ac{uc}.
    \item \textbf{Automated regularization, weights, and parameter estimation of ML software:} ML models in ML software often need to be fine-tuned with regularization, weights, and other parameters to obtain a more accurate convergence. Furthermore, ML models can get caught in local minima, and one of the best practices is to train multiple times to use the trained model with the optimum performance. In a machine learning software configuration with continuous training and automatic deployment, it is not always possible to perform this repeated training and fine tuning. In modern implementations, systematic parameter search looks for the best match among a predefined set of parameters by the user. Thus, there is potential for automated and principled parameter optimization.
    \item \textbf{Automated ML software container fault detection:} Failures can occur in containerized ML software running on edge and fog nodes. One of these flaws is that containers can suddenly shut down in the middle of a task. This can be caused by a variety of reasons, including an external attack. Currently, platforms such as Kubernetes provide a minimal container monitoring configuration based on the input of a network administrator \cite{kubernetes_tools_monitoring}. With the QA for ML, it is possible to implement automatic monitoring to discover and classify faults.
    \item \textbf{Drift detection and adaptation:} Streaming data is often subject to data and concept drift. The standard practice in this case is to retrain an ML software periodically or whenever a drift is found. An adaptation technique that retrains and redistributes weights on the fly without having to produce a new software version is a future direction in this research. An example of this is in the utilities sector with electricity data from smart meters. To manage electricity supply and demand, data from smart meters from individual and business users are used to estimate load forecasting and peak usage. When users install their solar panels or buy an electric vehicle, their normal pattern of electricity demand changes and drift occurs. As a result, with the help of QA for ML, drift adaptation can be implemented.
\end{itemize}

\section{Conclusion} \label{sec:conclusion}
In this paper, we presented an industrial perspective of QA in MLOps. We presented an architectural view of QA for ML and an architecture for industrial MLOps with automatic training and deployment of ML software. Furthermore, we identified some challenges and future directions of \ac{qa} in MLOps with industrial \ac{uc}s. This work will serve as a springboard for automated ML-based SQA and ML for the future factory and research inspiration.

\section*{Acknowledgments}
This work has been funded by the Knowledge Foundation of Sweden (KKS) through the Synergy Project AIDA - A Holistic AI-driven Networking and Processing Framework for Industrial IoT (Rek:20200067).

\bibliographystyle{unsrt}  
\bibliography{manuscript.bib}

\end{document}

%% file: acronyms.tex
\DeclareAcronym{qa}{
	short = QA,
	long = {Quality Assurance}
}

\DeclareAcronym{ml}{
	short = ML,
	long = {Machine Learning}
}

\DeclareAcronym{sqa}{
	short = SQA,
	long = {Software Quality Assurance}
}

\DeclareAcronym{iot}{
	short = IoT,
	long = {Internet of Things}
}

\DeclareAcronym{aws}{
	short = AWS,
	long = {Amazon Web Services}
}

\DeclareAcronym{uc}{
    short = UC,
    long = {Use Case}
}

\DeclareAcronym{dq}{
    short = DQ,
    long = {Data Quality}
}